\documentclass[aps,prl, twocolumn,secnumarabic,amsmath,amssymb]{revtex4}
\input epsf
\usepackage{graphicx}
\usepackage{url}

 \usepackage{setspace}
\batchmode

\begin{document}
\title{Optical bistability at low light level due to collective atomic recoil}
\author{M. Vengalattore}
\altaffiliation[email: ]{mukundv@calmail.berkeley.edu}
\author{M. Hafezi}
\author{M. D. Lukin}
\author{M. Prentiss}
\affiliation{MIT-Harvard Center for Ultracold Atoms \\
    Harvard University, Cambridge, MA - 02138}

\begin{abstract}
We demonstrate optical nonlinearities due to the interaction of weak
optical fields with the collective motion of a strongly dispersive
ultracold gas. The combination of a recoil-induced resonance (RIR)
in the high gain regime and optical waveguiding within the
dispersive medium enables us to achieve a collective atomic
cooperativity of $275 \pm 50$ even in the absence of a cavity. As a
result, we observe optical bistability at input powers as low as 20
pW. The present scheme allows for dynamic optical control of
the dispersive properties of the ultracold gas using very weak
pulses of light. The experimental observations are in good agreement
with a theoretical model.
\end{abstract}

\maketitle

Motivated by potential applications to quantum information science
\cite{imam1, lukin1, duan1, fleisch1}, there have been intense
experimental efforts to realize strong nonlinear interactions
between dilute atomic ensembles and weak optical fields. Methods to
achieve such quantum nonlinear couplings in dissipation-free media
have relied mainly on two approaches. First, the interaction time
and the coupling between the atoms and the photons can be enhanced
by placing the atoms within a high finesse cavity \cite{raimond}, an
approach that comes at the expense of considerable experimental
complexity and low bandwidth. Alternatively, near-resonant light
propagating through optically dense media can also result in strong
nonlinear interactions.  In order to limit the dominant linear
absorption, this latter approach requires the use of a coherent
multiphoton process such as EIT \cite{lukin2}. An added benefit of
such a multiphoton process is the enhancement of the interaction
time due to slow group velocities. However, since EIT relies on
quantum interference between the internal states of an atom, it is
fairly sensitive to inhomogenous optical and magnetic fields.

In this Letter, we demonstrate nonlinear optical effects due to the interaction between weak pulses
of light and the collective motion of an ultracold slow-light medium. Using the motional degrees of
freedom to create a highly dispersive gas alleviates the sensitivity to external fields.
As shown in recent studies \cite{veng2, veng1}, the implementation of a recoil-induced
resonance (RIR) in an optically dense anisotropic gas allows for strong atom-light interaction
due to the combination of slow group velocities and transverse confinement of the optical fields
within the atomic medium. Due to this strong coupling, we observe optical bistability at
input powers as low as 20 pW, an upper bound limited mainly by the photodetector efficiency.
\begin{figure}
\centering
\includegraphics[width=0.40\textwidth]{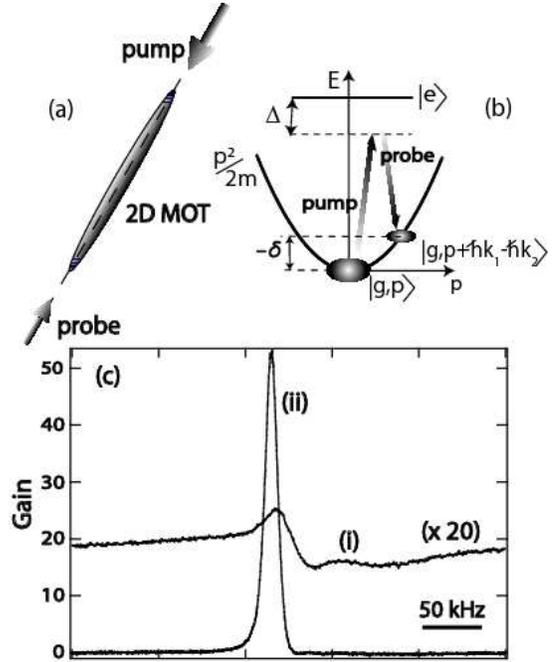}
\caption{(a) Schematic indicating the pump and probe beams
used for the RIR. (b)  The energy levels relevent to the RIR.
(c) Absorption spectrum of a probe beam
around the RIR for longitudinal optical densities of $\sim10$ (i)
and $\sim40$ (ii).}
 \label{fig:fig1}
\end{figure}

The motion of delocalized atoms under the influence of two or more light fields can mediate
the conversion of atomic kinetic energy into radiation \cite{courtois1}. These processes, termed
`recoil-induced resonances', can be described in terms of stimulated Raman transitions between
different momentum classes of the atomic ensemble \cite{guo1}. For an optically thin medium,
the atom-light interaction has little effect on the momentum distribution of the gas and a
perturbative analysis reveals that the probe field experiences weak absorption (gain) for positive
(negative) detuning relative to the pump field.

In contrast, as the optical density of the atomic gas increases, the
strong atom-light coupling leads to large optical gain \cite{veng2}
and significant effects on the momentum distribution of the atomic
gas. The strong amplification of the probe field and its subsequent
back-action on the collective motional states of the gain medium
results in a nonlinear optical response even for weak incident
beams. In this work, this back-action is evidenced by the
observation of optical bistability in the probe transmission.

\begin{figure}[t]
\centering
\includegraphics[width=0.40\textwidth]{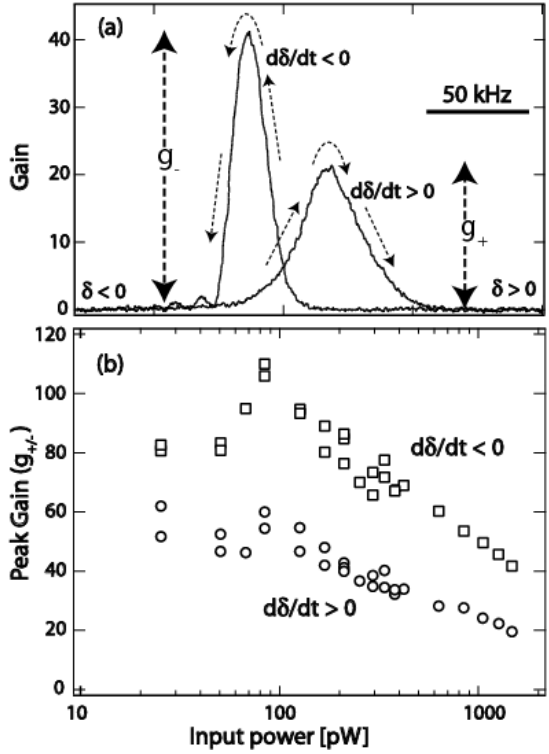}
  \caption{(a) Absorption bistability due to the interaction between the probe and the collective motion
of the atomic gas. Depending on the sign of the detuning chirp ($d\delta/dt$), the transmission indicates
a shift in both the resonance and the peak amplification ($g_+, g_-$). (b) Peak probe amplification indicates
a bistable response down to the detection limit of the input probe power.}
  \label{fig:fig2}
\end{figure}

The experimental scheme to create a strongly dispersive ultracold
gas using the RIR is described in previous work \cite{veng2}. About $5
\times 10^8$ $^{87}$Rb atoms are confined in a highly anisotropic
magneto-optic trap at typical temperatures of $\sim$ 20 $\mu$K. In
this trap, the atom cloud assumes the shape of a cylinder with
approximate radial (longitudinal) extent of 200 $\mu$m (3 cm). The
pump and probe beams for the RIR share the same linear polarization
and are directed along the long axis of this cylinder to take
advantage of the large optical depth (OD) along this axis. The Rabi
frequency of the pump beam was typically $\Omega_1/\Gamma = 1.5$
where $\Gamma$ is the natural linewidth of this transition. At low
OD, the transmission spectrum of the RIR exhibits a characteristic
dispersion-shaped spectrum with gain (absorption) for detuning
$\delta < 0$ ($\delta > 0$). At higher OD, the probe is almost
completely extinguished on the absorption side of the resonance
leaving the gain peak as the only distinguishable feature (Fig.\,1).

For an atomic gas with high OD, Fig.\,2(a) shows the probe
transmission as the pump-probe detuning is scanned across the RIR.
Depending on the sign of this detuning chirp, there a shift in the
resonance as well as in the maximum gain. This hysteretic nature of
the transmission diminishes as the OD is lowered to less than $\sim
10$. At larger cooperativity, the atomic ensemble and the light
fields form a strongly coupled system and the probe interacts with
an atomic ensemble whose motional coherences and momentum
distribution is the cumulative result of prior interactions with the
light fields. As seen in Fig.\,2(b), the bistable response ($g_-/g_+
> 1$) persists down to the detection limit ($\sim 20$ pW) of the
input probe power.

In order to understand this behavior, we first note that the
amplification of the probe is accompanied by the transfer of atoms
from a momentum $p$ to a momentum $p + 2 \hbar k$ (Fig.\,1(b)).
Thus, as the detuning is scanned across the gain side of the RIR,
atoms at various momenta are brought into resonance with the light
fields and transferred to higher momentum states. Crucially, in the
case of a negative chirp ($d \delta/dt < 0$), these transferred
atoms are brought {\em closer} to resonance with the light fields as
the detuning is scanned. Thus, in this process, atoms are
progressively swept up the momentum ladder due to the time-varying
detuning, resulting in both an enhanced amplification and a shift in
the location of the RIR. In contrast, for a positive chirp ($d
\delta/dt > 0$), the atoms are transferred to states that are {\em
farther} from resonance and the probe transmission resembles that
obtained for a static thermal distribution.

To quantitatively explain these observations, we use a theoretical model
that describes two classical light fields that are coupled to the
motional degrees of freedom of an elongated ensemble of two-level
atoms. To mimic the experiment, the atoms are assumed to be tightly
confined in the radial dimension. Accordingly, only the atomic
momentum along the long axis is relevent and the Hamiltonian can be
written as \cite{moore98}
\begin{eqnarray}
\mathcal{H} &=& \sum_k [ \frac{\hbar^2 k^2}{2 m}c_{g}(k)^{\dagger}c_{g}(k) + (\frac{\hbar^2 k^2}{2 m}
+ \hbar \omega_0) c_{e}(k)^\dagger c_{e}(k) \nonumber \\
&& + i \hbar \sum_{j=1,2} (g_j a_j^\ast e^{i \omega_j t} c_{g}(k-k_j)^\dagger c_{e}(k) - h.c.) ]
\end{eqnarray}
where $c_{g}(k) (c_{e}(k))$ are the annihilation operators of ground (excited) state atoms with momentum
$\hbar k$, $\omega_0$ is the transition frequency of the two-level atoms, $g_ 1(g_2)$ is the atom-light coupling
coefficient and $a_{1} (a_2)$ is the normalized electric field of the pump (probe) beams. In the far-detuned
limit, the excited states can be adiabatically eliminated and the equation for the coherences and populations
of the different momentum classes in the ground state are
\begin{eqnarray}
&&\frac{d}{dt}\rho(p,p') = 4 i \omega_r (p'^2 - p^2)\rho(p,p') \nonumber \\
&+& i \frac{g_1 g_2 a_1^\ast}{\Delta}a_2 e^{-i\delta t}(\rho(p+1,p') - \rho(p,p'-1))\nonumber \\
&+& i \frac{g_1 g_2 a_1}{\Delta} a_2^\ast e^{+i \delta
t}(-\rho(p,p'+1) + \rho(p-1,p'))
\end{eqnarray}
where $\rho(p,p') \equiv \langle c^{\dagger}_g(k') c_g(k) \rangle$,  $\omega_r$ is the recoil frequency and
$\delta = \omega_2 - \omega_1$  is the detuning. Also, using the slowly-varying envelope and single-mode
approximations, the evolution of the probe amplitude can be written as
\begin{equation}
\frac{d}{dt}a_2 = i N \frac{g_1 g_2}{\Delta} a_1 e^{i \delta t}
\sum_p \rho(p-1,p) - \frac{\kappa}{2} (a_2 - a_{in}).
\end{equation}

Retaining only first-order coherence terms between momentum classes, the above equations can be written
as a set of coupled equations for the population $\Pi_p=\rho(p,p)$, the first-order coherence $\eta_p \equiv \rho(p+1,p)
e^{i \delta t}$ and the probe amplitude $a_2$.
\begin{eqnarray}
\dot{\Pi}_p &=& \left[-i \beta^\ast a_2 ( -\eta_p+\eta_{p-1} ) + c.c.\right] - \gamma_{pop} (\Pi_p-\Pi_{th, p}) \nonumber \\
\dot{\eta}_p &=& i(4 \omega_r (p^2 - (p+1)^2) - \delta(t) + i\gamma_{coh})\eta_p  \nonumber \\
&-& i \beta a_2^\ast (\Pi_{p+1}-\Pi_p) \nonumber \\
\dot{a}_2 &=&  i \beta N \sum_p \eta^\ast_{p-1} - \frac{\kappa}{2} (a_2 - a_{in})
\end{eqnarray}
where $\gamma_{pop} (\gamma_{coh})$ are the population (coherence) relaxation rates respectively and
$\beta = g_1 g_2 a_1/\Delta$. The thermal population distribution $\Pi_{th, p}$ is given by Maxwell-Boltzman distribution.  The decay rate of photons is approximated by the free-space rate $\kappa = c/L$
with $L$ the longitudinal extent of the atomic gas. These coupled equations describe the rich dynamics that
ensue as a consequence of the collective atom-light interaction and a time-dependent pump-probe detuning.

Numerical simulations of the bistability based on this model show excellent agreement with the experimental
results over a wide range of parameters of pump detuning and chirp rate (Fig. 3). In these simulations, parameters
such as the pump detuning, OD of the atomic gas and scan rate of the two-photon detuning were held fixed at
the experimental values.
\begin{figure}[h]
\centering
\includegraphics[width=0.50\textwidth]{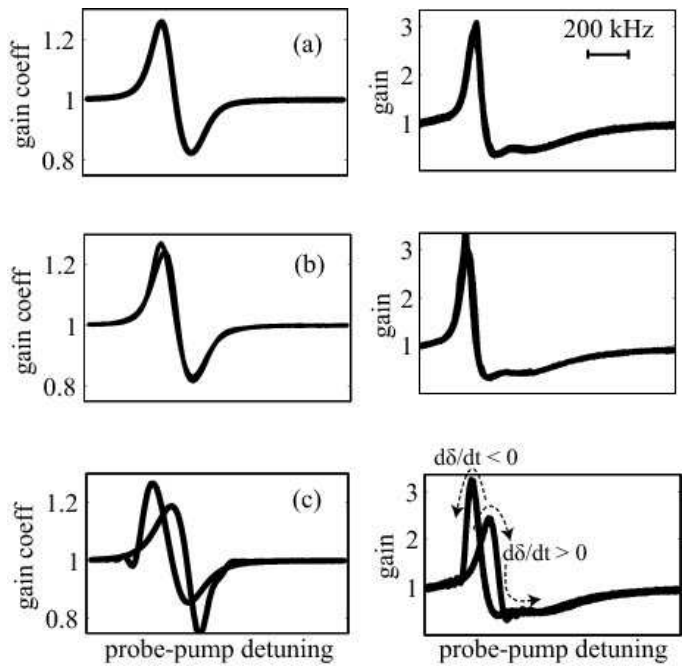}
\caption{Numerical simulations based on the theoretical model match
the experimental results over a wide range of parameters. Panels on
the left show the numerical simulations for the gain coefficient
given by $exp(-2 Re[\alpha]L)$ where $\alpha$ is the absorption
coefficient and panels on the right show the observed experimental
gain of the probe transmission across the RIR, both versus
two-photon detuning.(a), (b) and (c) correspond to transmission spectra obtained
by chirping the two-photon detuning at scan rates of
0.1, 0.5 and 2.5 MHz/ms, respectively.} \label{fig:match}
\end{figure}

The observation of a bistable transmission requires that the momentum coherences established as a result of
the atom-light interaction persist as the detuning is scanned across the RIR. In practice, off-resonant light scattering
causes the thermalization of the momentum distribution thereby suppressing the bistability. This competing
process leads to a limiting rate for the scan rate ($d \delta /dt$) below which the atomic momentum distribution
is quasi-static and the transmission becomes non-hysteretic (Fig.\,4(a)). A bistable transmission was observed
for scan rates as low as 500 kHz/ms, corresponding to moving across the RIR in $100 \, \mu$s, much longer
than the mean photon scattering time of $<\, 1\, \mu$s. This observation of long-lived momentum coherences
is consistent with previous studies of the RIR \cite{kozuma, courtois1}.


\begin{figure}[b]
\centering
\includegraphics[width=0.48\textwidth]{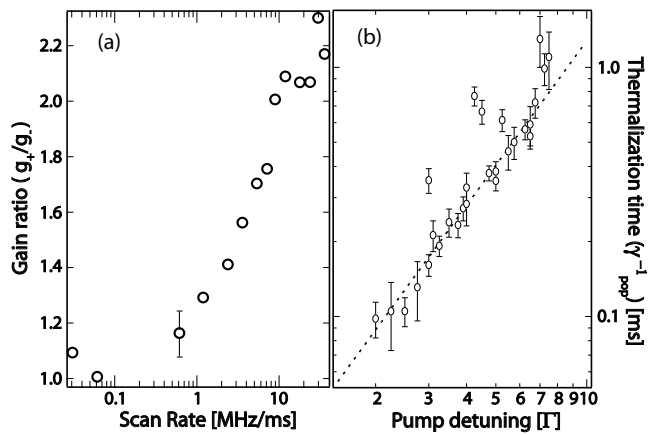}
  \caption{(a) The ratio of peak gain ($g_-/g_+$) indicates a limiting scan rate $d \delta/dt \approx 0.5$ MHz/ms
below which decoherence and thermalization of the momentum distribution suppress optical bistability.
This data was obtained at a pump detuning $\Delta \sim -4 \, \Gamma$.
(b) Thermalization time ($\gamma_{pop}^{-1}$) {\em vs} the pump detuning. A fit to the data (dashed
line) indicates a power-law dependence $\gamma_{pop}^{-1} \propto \Delta^\alpha$ with $\alpha = 1.57
\pm 0.09$.}
\label{fig:fig3}
\end{figure}

An independent measure of the influence of off-resonant scattering
on the probe transmission was obtained by determining the
thermalization time ($\gamma_{pop}^{-1}$) of the atomic momentum
distribution. For this, the probe was switched on for 100 ms at an
intensity of 0.1 mW/cm$^2$ and at a two-photon detuning that
corresponded to the peak gain of the RIR. The probe intensity was
then reduced to $\sim 10^{-3}$ mW/cm$^2$ within a few microseconds.
Following this reduction in intensity, the probe transmission
indicated a gain that was initially very small but gradually relaxed
to a higher equilibrium value. This relaxation was interpreted as
being due to the thermalization of the momentum distribution to
repopulate those states that were depleted due to the initially
intense probe.  Under the assumption that the final probe intensity
was too small to significantly modify the atomic distribution, the
time scale over which the probe transmission reaches a steady state
should reflect the thermalization time. Consistent with the
observation of bistability at relatively slow scan rates, we measure
thermalization times on the order of a few hundred microseconds
(Fig.\,4(b)).

The numerical simulations used the measured value of $\gamma_{pop}$
(Fig.\,4(b)) while allowing the value of $\gamma_{coh}$ to vary. In
order to obtain the best match with the experimental results, we
typically found that $\gamma_{coh}$ was required to be greater than
$\gamma_{pop}$. However, both quantities were much smaller than the
photon scattering rate. This suppression is in agreement with
earlier observations \cite{verkerk92,courtois1,kozuma,courtois2}.
A possible explanation lies in the fact that in the presence of an optical
potential ($U \propto 1/\Delta$), in the limit of large detuning,
both $\gamma_{coh}$ and $\gamma_{pop}$ can be suppressed due to
Lamb-Dicke confinement. The estimated scaling of the suppressed
decay rate with the detuning should be $\gamma_{eff}\propto
\gamma_{scatt}/\sqrt{U} \propto \Delta^{-3/2}$ where
$\gamma_{scatt}$ is the  photon scattering rate. The experimental
data (Fig.\,4(b)) also suggests a similar scaling($1.57\pm0.09$).


The observation of optical bistability at low input powers suggests prospects of all-optical control using
weak pulses of light. For instance, this scheme lends itself rather easily to a low light level all-optical switch
wherein the detuning of a few-photon probe is controlled in order to switch a more intense output beam.  The
largest scan rates, $d \delta/dt = 30$ MHz/ms at which bistability was observed and typical values of the
momentum coherence time $\gamma^{-1}_{coh} \sim 100 \, \mu$s together yield an estimate of the
switching time $\tau =  \gamma_{coh}/(d \delta/dt) =$ 0.3 $\mu$s. This is commensurate with previous
measurements of all-optical switching using a RIR \cite{veng2}. Combining this with the lowest probe powers
($P \sim 20$ pW) and typical beam waists (100 $\mu$m) used in this work, we obtain a typical photon number
$\tau P/(h c/\lambda) \sim 25$ and a remarkably low switching energy density \cite{gauthier1,zhang07} of $7\times 10^{-5}$
photons/($\lambda^2/2 \pi$) to operate this all-optical switch.

In conclusion, we demonstrate optical nonlinearities due to a
coherent interaction between weak light fields and the collective
motion of a strongly dispersive atomic gas. Since the atomic
momentum is relatively insensitive to external magnetic or electric
fields, such systems may be promising candidates for applications in
low-light level nonlinear optics.

This work was funded by the NSF, the Center for Ultracold Atoms,
DARPA and Packard foundation.

\end{document}